\documentclass[lettersize,journal]{IEEEtran}
\usepackage{multirow}
\usepackage{amsfonts,amsmath,amssymb,mathtools}
\usepackage{graphicx}
\usepackage{epstopdf}
\usepackage{xurl}
\usepackage{cite}
\usepackage[caption=false,font=footnotesize,subrefformat=parens,labelformat=parens]{subfig}
\usepackage{comment}
\usepackage[ruled,vlined]{algorithm2e}
\usepackage{algpseudocode}
\usepackage{tikz}
\usetikzlibrary{arrows,intersections,decorations,plotmarks,math,patterns}
\usepackage{pgfplots}
\pgfplotsset{compat=1.17}
\usepackage{color}
\usepackage[colorinlistoftodos]{todonotes}

\usepackage{hyperref}

\definecolor{COL11}{RGB}{215,25,28}
\definecolor{COL12}{RGB}{253,174,97}
\definecolor{COL13}{RGB}{146,197,222}
\def\BibTeX{{\rm B\kern-.05em{\sc i\kern-.025em b}\kern-.08em
    T\kern-.1667em\lower.7ex\hbox{E}\kern-.125emX}}
\usepackage{balance}
\usepackage{booktabs}
\usepackage[scr=rsfs]{mathalpha}
\usepackage{makecell}
\usepackage[export]{adjustbox}
\usepackage{autobreak}

\begin{document}

\title{ Lightweight Ac Arc Fault Diagnosis via Fourier Transform Inspired Multi-frequency Neural Network}

\author{
Qianchao Wang$^{1}$ , Chuanzhen Jia$^{2}$, Yuxuan Ding$^{1\dag}$, Zhe Li $^{3}$, Yaping Du $^{1}$

\thanks{Yuxuan Ding is the corresponding author}
\thanks{$^{1}$ Qianchao Wang, Yuxuan Ding, and Yaping Du are with the Department of Building Environment and Energy Engineering, Hong Kong Polytechnic University, Hung Hom, Hong Kong. (qianchao.wang@polyu.edu.hk; chuanzhen.jia@connect.polyu.hk; yx.ding@connect.polyu.hk; Ya-ping.du@polyu.edu.hk )}
\thanks{$^{2}$ Chuanzhen Jia is with the Department of Electrical and Electronic Engineering, Southern University of Science and Technology, Shenzhen 518055, China and also with the Department of Building Environment and Energy Engineering, The Hong Kong Polytechnic University, Kowloon, Hong Kong, China. (e-mail: chuanzhen.jia@connect.polyu.hk).}
\thanks{$^{3}$ Zhe Li is with Shenzhen Power Supply Bureau Co., Ltd, Shenzhen, China. (19109887r@connect.polyu.hk)}
\thanks{This work was supported by the Science and Technology Project of China Southern Power Grid Co., Ltd. ``Research on Data Set Platform and Testing Standard for Low Voltage Electricity Safety Hazar''(090000KC24110001).} 

}

\maketitle

\begin{abstract}
 
Lightweight online detection of series arc faults is critically needed in residential and industrial power systems to prevent electrical fires. Existing diagnostic methods struggle to achieve both rapid response and robust accuracy under resource-constrained conditions.
To overcome the challenge, this work suggests leveraging a multi-frequency neural network named MFNN, embedding prior physical knowledge into the network. Inspired by arcing current curve and the Fourier decomposition analysis, we create an adaptive activation function with super-expressiveness, termed EAS, and a novel network architecture with branch networks to help MFNN extract features with multiple frequencies.
In our experiments, eight advanced arc fault diagnosis models across an experimental dataset with multiple sampling times and multi-level noise are used to demonstrate the superiority of MFNN. The corresponding experiments show: 1) The MFNN outperforms other models in arc fault location, befitting from signal decomposition of branch networks. 2) The noise immunity of MFNN is much better than that of other models, achieving 14.51\% over LCNN and 16.3\% over BLS in test accuracy when SNR=-9. 3) EAS and the network architecture contribute to the excellent performance of MFNN.

\end{abstract}


\section{Introduction}\label{Introduction}

\IEEEPARstart{W}{ith} the increasing electrification of buildings and transportation systems, arc faults have become a critical concern in power systems. These faults can ignite surrounding combustible materials, posing significant risks to both human safety and property~\cite{thakur2024advancementsarcfaultdetection}. Meanwhile, arc fault often occurs in the high-resistance state in the initial stage, rendering conventional over-current and leakage current protection devices ineffective in fault detection~\cite{jiang2021coupling}. Therefore, in addition to physical-signal methods, recent research addresses this challenge by embedding machine learning methods into arc fault diagnosis~\cite{liu2024arc}. To ensure the equipment deployment and online detection of arc fault diagnosis, lightweight models are highly concerned.


These methods can be divided into two categories: feature extraction-based methods and end-to-end deep learning methods. Feature extraction-based methods analyze signals in the time and frequency domains, using empirical wavelet transform~\cite{yi2022arc}, wavelet transform~\cite{9950518}, and Fourier Transform~\cite{8624391} to derive spectral features and frequency information, enabling effective diagnosis in simple cases~\cite{jiang2021coupling}, especially with machine learning~\cite{8970332} or deep learning models~\cite{10141554}. Nevertheless, these methods struggle with loads like cleaners and dimmers, where non-arcing signals span a wide frequency range, resembling arcing signals~\cite{9720225}. Their two-stage process is computationally intensive and less robust, as performance depends on accurate feature extraction and artificial stage coupling.

To address this issue, end-to-end methods, for instance, deep learning, are used as another way for arc fault diagnosis. It can effectively avoid the inconvenience caused by the two-stage methods and extract the necessary features adaptively~\cite{8571267,wang2024interpretable}.
The raw or pre-processed electrical signals (current or voltage) are directly leveraged as the input of deep neural networks, using the powerful feature extraction ability of networks to adaptively capture useful features. Some novel networks have demonstrated their out-performance in arc fault diagnosis, such as AutoEncoder (AE)~\cite{9439848}, series arc fault neural network (LCNN)~\cite{10018466}, ArcNet~\cite{9392282}, and temporal convolution network (ArcNN)~\cite{10054597}. 
{They use CNN to extract necessary features to improve the fault diagnosis accuracy. Meanwhile, some advanced general models or hybrid models are utilized in arc fault diagnosis, such as Transformer-based classifiers~\cite{chabert2023transformer}, Multimodal models~\cite{11007029}, CNN-LSTM~\cite{10908361}, and broad learning system (BLS)~\cite{10477286}. These methods are normally utilized to solve more challenges, for example domain shifting and data diversity.}
The high fault diagnosis accuracy implies that these end-to-end methods can automatically discover the potential necessary features without manual screening or a wide frequency range search. Whereas, the methods are also computationally intensive, which is the same as the feature-based methods, since the feature extraction plays an equally important role in both methods. Meanwhile, without physical guidance, the trustworthiness of these highly accurate models is questionable, due to the black-box nature of the network~\cite{samek2019explainable}. 

Therefore, the development of arc fault diagnosis faces two challenges: 1) first, the computationally intensive methods are limited by computational resources. 2) second, black-box methods without physical guidance are questionable. In addressing the two bottlenecks, a core obstacle is that \textit{`In limited computational resources, can we guarantee the arc fault diagnosis accuracy using physical guidance in deep learning?'}. This question can be answered from two aspects: \textit{`what physical guidance can be used?'} and \textit{`how to map physical information to deep learning?'}. By analyzing the electrical signals of the arc, the signals can be recognized as the combination of multiple signals with different periods~\cite{10130787}, in which the high harmonics play a paramount role to achieve precise arc fault detection~\cite{liu2024arc}. As long as we find high harmonics as key features, the arc fault diagnosis model can perform well. Therefore, physical guidance becomes two things: periodic signals and signal decomposition. The key problem becomes the mutual mapping between physical guidance (the periodic signals and signal decomposition) and deep learning models.

In this vein, the primary objective of this work is to design a novel neural network equipped with physical guidance by mapping the periodic signals and signal decomposition into the network components, to adaptively find the fault signal frequency, guaranteeing the arc fault diagnosis accuracy in limited computational resources. In this paper, we propose a new deep learning model for arc fault diagnosis, termed multi-frequency neural network (MFNN), by creating a new activation function and model structure. Inspired by~\cite{sitzmann2020implicit}, the new activation function has `Sin' with adaptive training parameters as the periodic positive part and an asymptotic function as the analytical negative part, ensuring the excellent super-expressiveness~\cite{yarotsky2021phase} and adaptive frequency extractions. Simultaneously, by analyzing the core idea of the Discrete Fourier Series (DFS) and embedding the branch networks in MFNN, an additive network structure can elegantly and efficiently help the deep learning model decompose the input signals and adaptively capture essential features with fewer parameters. 
Notably, this novel neural network holds the potential for adaptation to other domains with similar prior knowledge in industrial diagnostic applications. To demonstrate the efficiency of MFNN, seven additional deep learning-based arc fault diagnosis models are leveraged for comparisons across arc datasets with different sampling times and signal-to-noise ratio (SNR). The ablation experiments verify that each physics-guided network component contributes to the good performance of MFNN, confirming the effectiveness of these components. In summary, the key contributions of this paper encompass: 
\begin{enumerate} 
	\item A novel unified neural network equipped with physical guidance, named MFNN, is proposed to deal with arc fault diagnosis in limited computational resources. 
    \item An adaptive activation function with super-expressiveness is proposed by setting a periodic positive part with trainable parameters and an analytical negative part with an asymptotic function to ensure the network expression capability under limited resources.
    \item By analyzing the core idea of DFS, a new neural network architecture is proposed by embedding several branch networks into MFNN to decompose the input signals and capture essential features adaptively.
    \item An arc fault experiment using a standard arc fault test platform is conducted. Meanwhile, eight advanced deep learning-based arc fault diagnosis models are tested across datasets with different sampling times and SNR to validate the effectiveness of MFNN. Each physics-guided component is tested in comparative and ablation experiments to analyze the contributions to MFNN. Hardware Implementation of MFNN is also tested
\end{enumerate}


The rest of this paper is organized as follows: Section~\ref{sec:Methodology} provides the background of the utilized methods. Section~\ref{sec:Ourmethods} explains the proposed MFNN and the built-in physic-guided components. Then, Section~\ref{sec:Experiments} shows all the experiments, including the test accuracy and the corresponding ablation experiments, and Section~\ref{sec:Conclusion} concludes the paper.


\section{PRELIMINARIES}
\label{sec:Methodology}

\subsection{Discrete Fourier Series and Its Physical Guidance}
This subsection introduces the latent physical guidance 
by analyzing the core idea of DFS when dealing with limited input signals. 

We assume that the given sampling input vector is $\mathbf{x}=[x_0,x_1,\dots,x_{N-1}], x_i=f(i\Delta t)$ with finite duration T, where $f:[0,T]\rightarrow \mathbb{C} $ is the mapping of $x_i$ and $\Delta t>0, i=0,1,\dots,N-1,T=N\Delta t$. By extending the input vector $\mathbf{x}$, 
there is a T-periodic sampling sequence $\mathbf{x_p}=[\dots,x_{-2},x_{-1},x_0,x_1,\dots,x_{N-1},x_{N},\dots]$. Then, the DFS converts the finite sequence of equally spaced samples of a function into a sequence of complex numbers representing the frequency components, which is given by
\begin{equation} \label{eq:residual}
\mathbf{X_k}=\sum_{i=0}^{N-1}f(i\Delta t)e^{-jkiw_0\Delta t}\Delta t= \sum_{i=0}^{N-1}x_ie^{-jki2\pi/N} 
\end{equation}
where $\mathbf{X_k}$ is the k-th frequency-domain signal and $k=0,1,2,\dots,N-1$. The frequency domain signal $\mathbf{X}=[X_0,X_1,\dots,X_{N-1}]$. Meanwhile, the time-domain input signal can be represented by the frequency-domain signal in Eq.~\ref{eq:residual} as well, which is given by
\begin{equation} \label{eq:DFS}
x_i= \sum_{k=0}^{N-1}\mathbf{X_k}e^{-jki2\pi/N} 
\end{equation}


The equation means that the input signal can be recognized as a linear combination of periodic signals with different frequencies $\mathbf{x}=\sum Signal(w_i)$, indicating the decomposability and additivity of the input signals. 


This kind of underlying physical guidance (decomposability and additivity) enforces that the structure of a neural network should be like a hydrological structure, with information converging from multiple branch rivers to the sea. This kind of structure has been mathematically proven to have enhanced expressive power when using fixed-size ReLU networks~\cite{zhang2023enhancingexpressivepowercompositions}.
From another perspective, the input signal can be viewed as a complex mathematical function that can be affected by the nonlinearity of multiple sub-inputs, and the generalized additivity model of the tree structure improves the inherent intelligibility of inputs~\cite{NEURIPS2021_251bd044}, helping models find the key features. Combined with the expressivity of deep neural networks, the models with the new structure can learn arbitrarily complex input signals, as long as physical guidance exists.

\subsection{Super-expressiveness of Activation Functions and Examples}
Super-expressiveness is based on the approximation ability of neural networks. From the perspective of approximation theory, a neural network employing activation functions with periodic and analytic components can approximate any continuous function with an arbitrarily small error, even with a fixed number of neurons~\cite{zhang2022deep,yarotsky2021phase}. This remarkable property is referred to as the super-expressiveness of such activation functions.

We assume that $\mathcal{NN_{\varrho}} \left \{ N,L \right \} $ is a $\varrho$-activated  neural network limited to a width of at most $N$ and a depth of at most $L$. The mapping of $\mathcal{NN_{\varrho}}$ is $\mathbb{R}^d \rightarrow \mathbb{R}^n$. The activation function $\varrho:\mathbb{R} \rightarrow \mathbb{R}$ satisfies the following conditions.
\begin{itemize}
     \item There exists an interval $(\alpha,\beta)$ with $\alpha<\beta$ such that $\varrho$ is real analytic and  non-polynomial on $(\alpha,\beta)$.
     \item There exists a fixed-size $\varrho$-activated network $\mathcal{NN_{\varrho}}$ reproducing a triangle-wave function on $[0,\infty)$.
     \item The activation functions $\varrho$ can be extended in the ``closure", $\bar{\mathscr{A}}$, of all similar activation functions.
\end{itemize}

\begin{figure}[ht] 
\centering
\subfloat{
\includegraphics[width=0.395\linewidth]{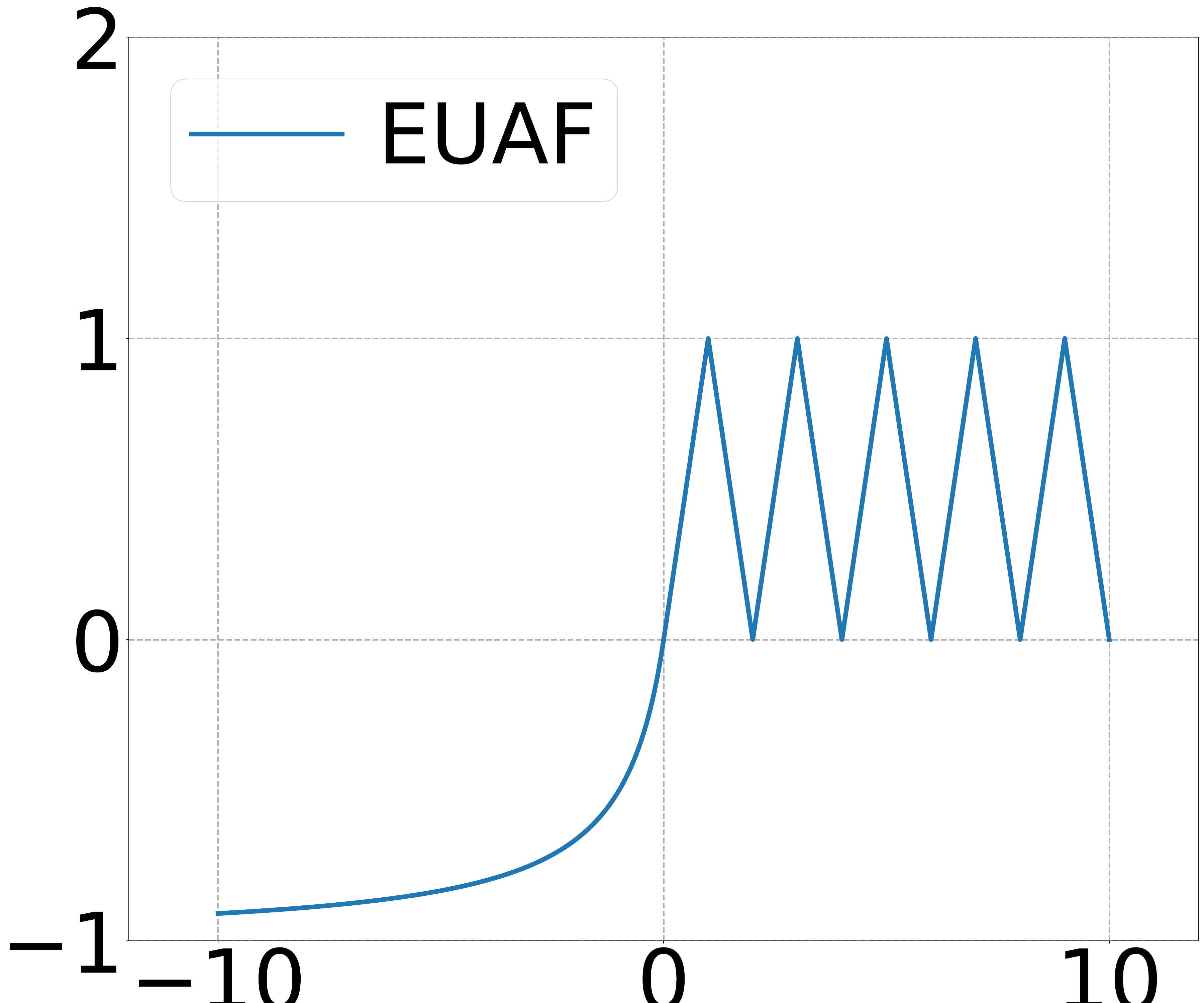}
}~\hspace{0.3cm}
\subfloat{
{\includegraphics[width=0.4\linewidth]{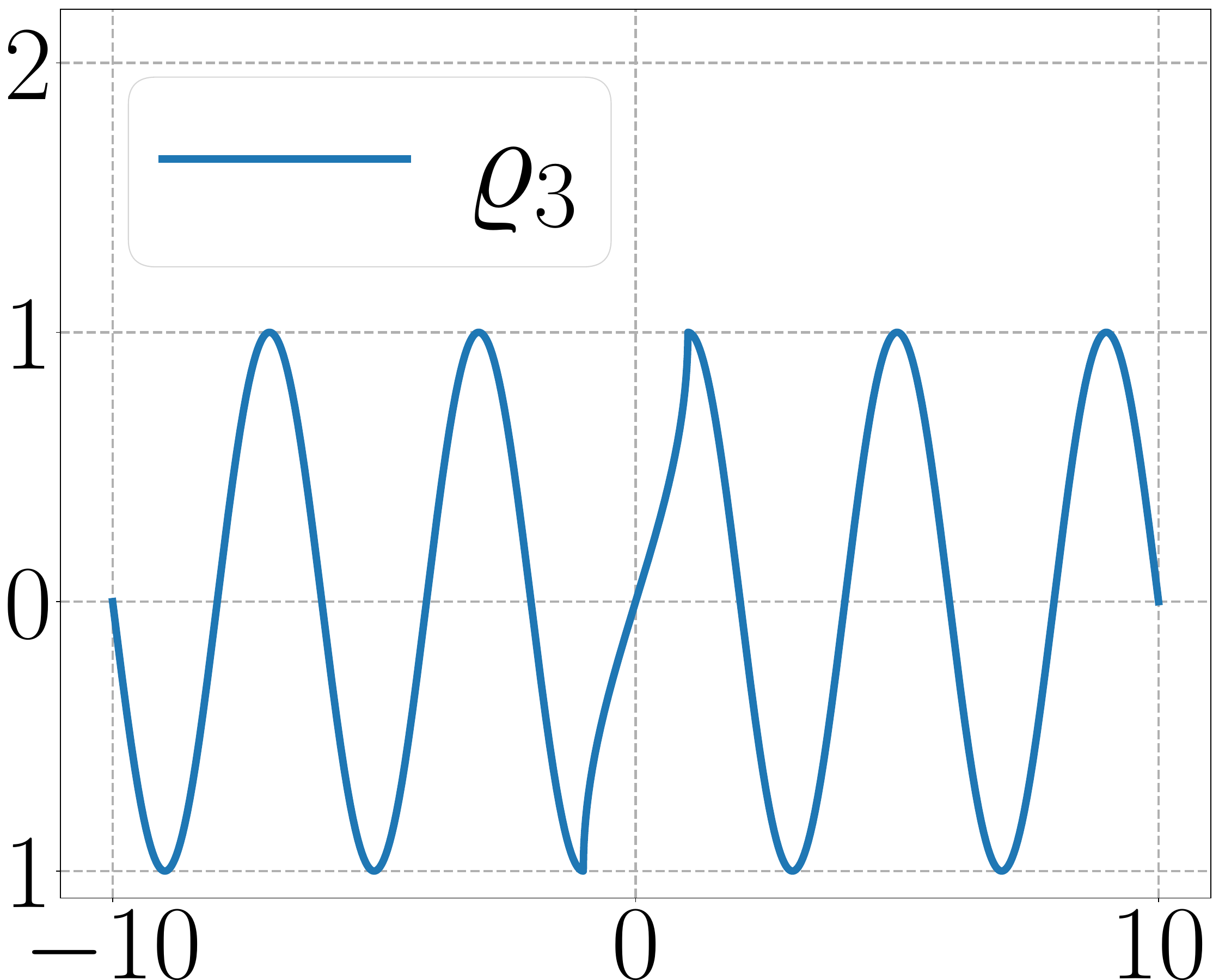}}
}
\caption{Illustrations of EUAF and SinTrx. }
\label{fig:super-expressiveness AFs}
\vspace{-0.5cm}
\end{figure}

We give two classic examples to visually describe the above conditions, depicted in Fig~\ref{fig:super-expressiveness AFs}. The first is the elementary universal activation function (EUAF) which follows the first two conditions, defined as:
\begin{equation*}
	\mathrm{EUAF}(x)\coloneqq \begin{cases}
		\big|x-2\lfloor \tfrac{x+1}{2}\rfloor\big|  & \text{for} \  x\ge  0,\\
		\frac{x}{1+|x|} & \text{for} \  x<0,\\
	\end{cases}
\end{equation*}
EUAF has an analytical function on $(-\infty,0)$ and a periodic function on $[0,\infty)$. The second is a sin-based activation function that follows the third condition, defined as:
\begin{equation*}
	\mathrm{SinTrx}(x)\coloneqq \begin{cases}
        \tfrac{2}{\pi}\arcsin(x) & \text{for}\ -1\le  x\le 1,\\
        \sin(\tfrac{\pi}{2}x) &\text{for}\  |x|>1. \\
	\end{cases}
\end{equation*}
SinTrx is periodic on $[1,\infty)$ and $(-\infty,-1)$. These unique activation functions allow neural networks to achieve precise approximation accuracy using the desirable property of super-expressiveness.

\subsection{Motivation with Data Analysis}~\label{Motivations}

Our goal is to design an effective arc fault diagnosis model that both embeds the physical constraints or guidance and unleashes the flexibility and expressivity of neural networks, improving the computational efficiency in limited computational resources.

\begin{figure}[!t] 
\centering
\includegraphics[width=0.85\linewidth]{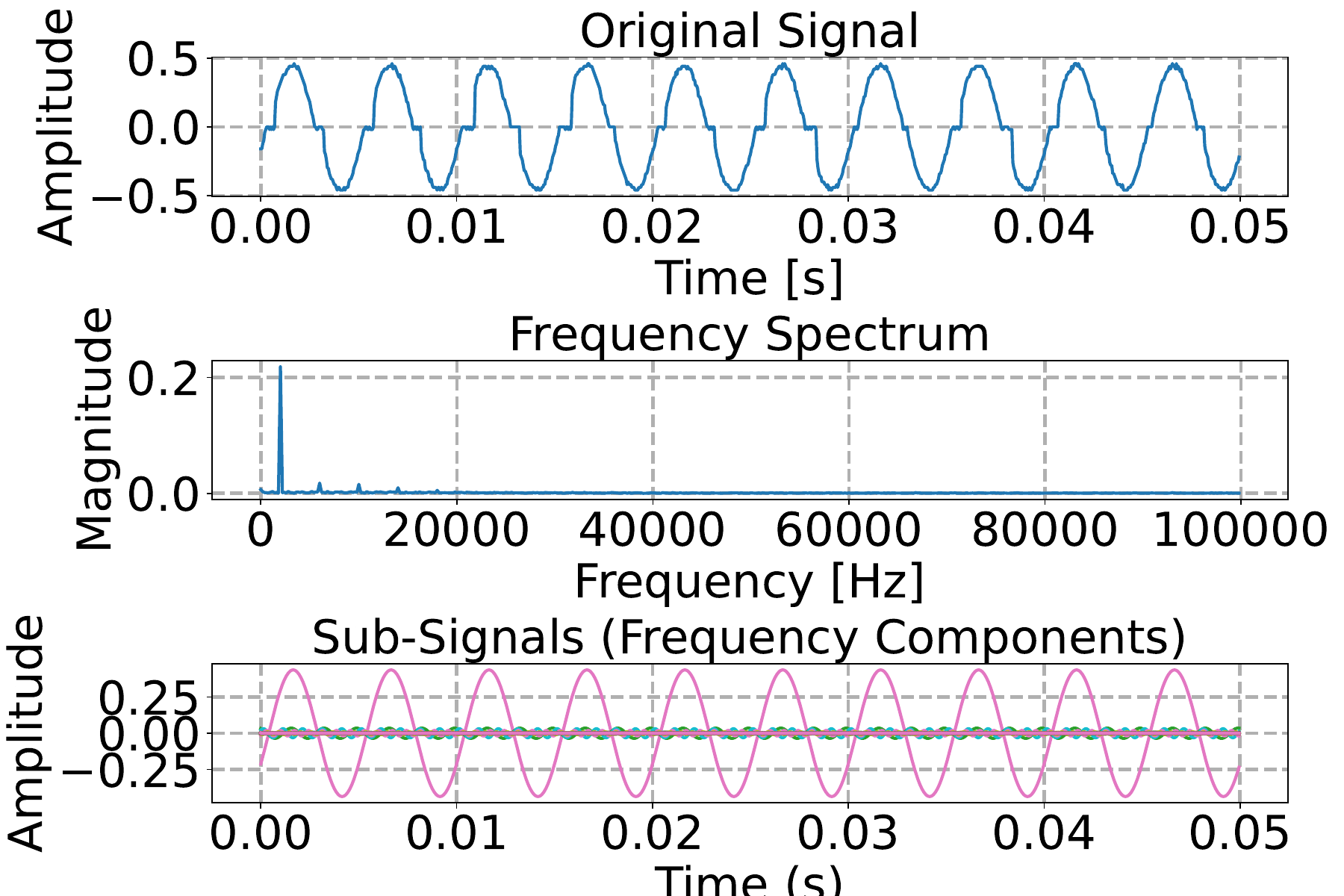}
\caption{The decomposition of an arc fault signal.} 
\label{fig:decomposition of arc}
\vspace{-0.3cm}
\end{figure}

Fig~\ref{fig:decomposition of arc} depicts the decomposition of an arc fault signal with resistive loads using the Fourier transform. As resistive loads are related to heat transfer, there are almost no harmonic components in synchronized current and voltage signals, which is reflected in the time-domain signal at the top of the figure. Meanwhile, the ``Flat Shoulder'' across the zero-crossing region in the arcing current is a key feature in the resistive load-related fault current. There are no significant spikes in the current signal. In general, fault signals are the distortion product of normal signals.

Nonetheless, these easily observable distortions are difficult to represent mathematically in the time domain, especially when datasets are described in tabular form. It prevents the signal from being recognized by the machine. However, the frequency-domain signals can compensate for this shortcoming at the cost of some memory. The magnitudes of the sub-signal at each frequency, shown in the middle sub-figure, indicate the proportion of sub-signals in the total signal. Except for the sub-signals under the main amplitude, other signals are potential fault features for arc fault diagnosis and are easier to understand by machines than time domain signals.
The bottom sub-figure gives the decomposed sub-signals, which indicate that the time domain signal is a linear coupling of multiple periodic signals with different frequencies and phases.

The intrinsic signal characteristics and the distortions mined by the Fourier transform suggest a promising avenue: leveraging the physical constraints or guidance could help neural network models locate the arc faults.

\section{Multi-frequency Neural Network}
\label{sec:Ourmethods}

In industrial model applications, trustworthiness, computational resources, and fault diagnosis accuracy are critical to solving a certain type of problem. Although scaling law theory has been widely acknowledged in computer science, the blind pursuit of large models is debatable in industry. In this section, we propose the multi-frequency neural network, embedding the physical guidance inspired by signal inherent characteristics and discrete Fourier transform in the network, to discuss how physics-inspired small models improve efficiency and robustness.

\subsection{Enhanced Adaptive Sine: A Physics-inspired Activation Function}

\begin{figure}[!t]
\centering
\subfloat[An illustration of EAS, where $\omega$ is 3 and $\varphi$ is $\frac{\pi}{4}$.]{
\includegraphics[width=0.45\linewidth]{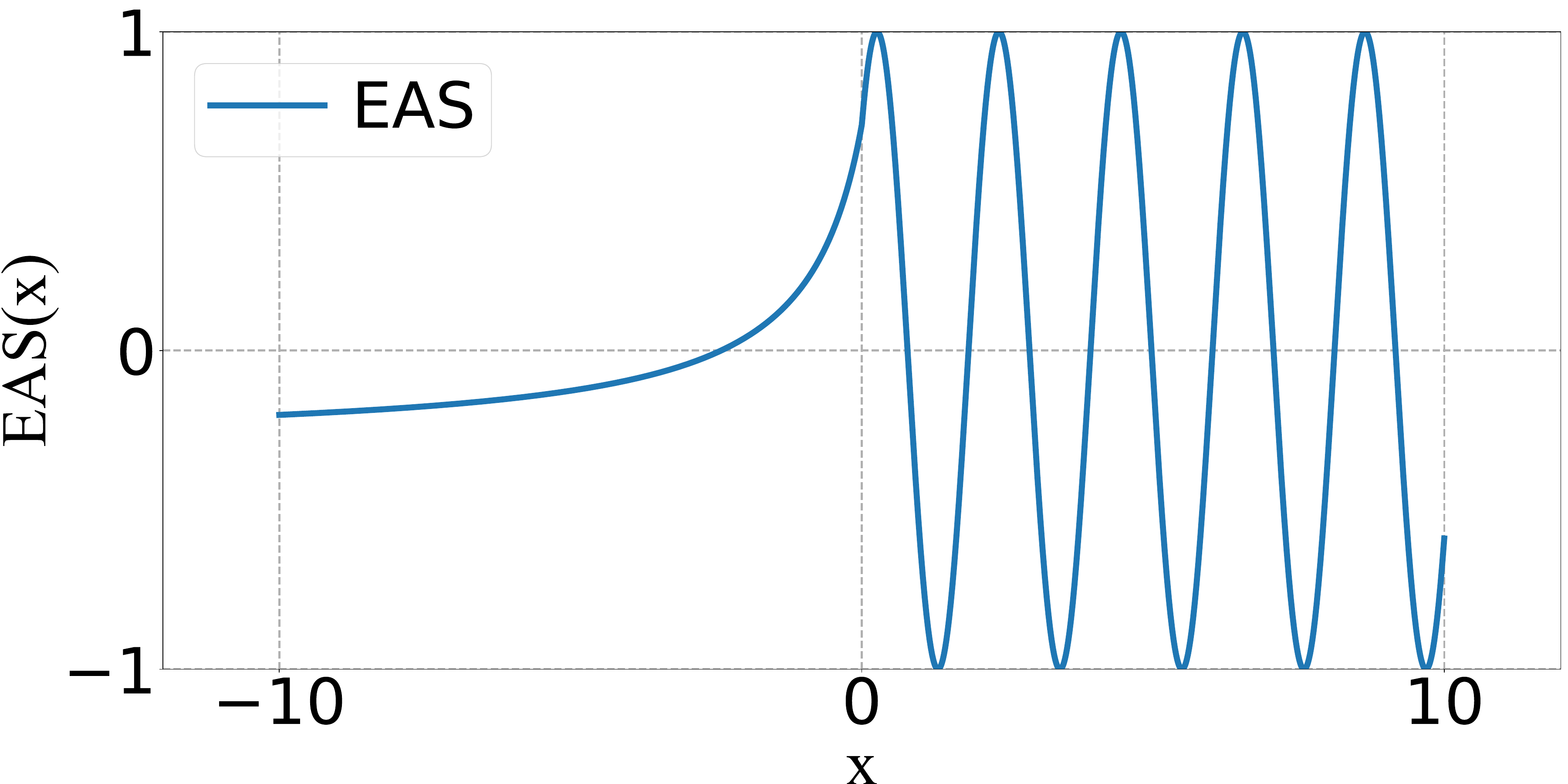}
\label{fig:An illustration of EAS}
} ~
\subfloat[The convergence of MFNN.]{
\includegraphics[width=0.45\linewidth]{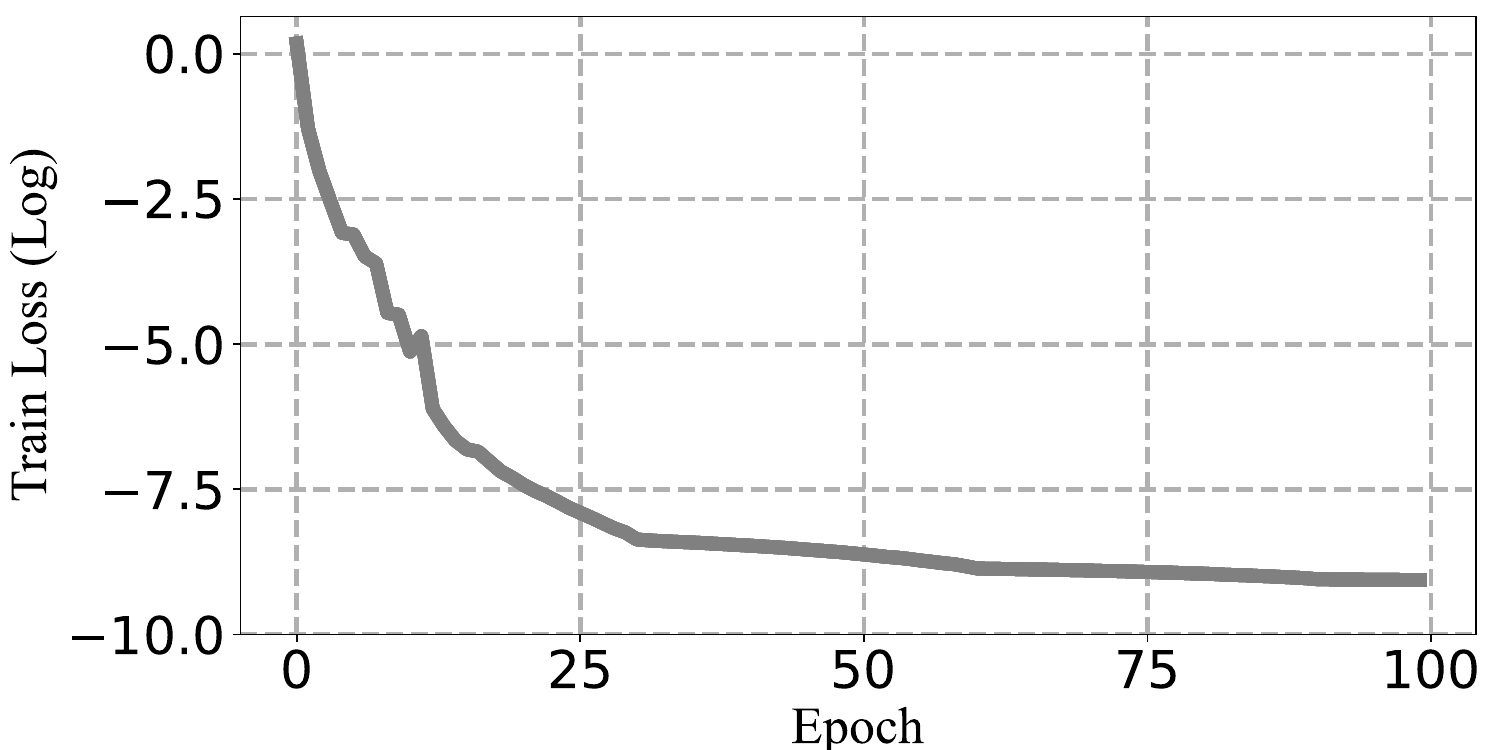}
\label{fig:The convergence of MFNN}
} \\
\caption{EAS example and the convergence of MFNN. }
\label{fig:EAS and convergence}
\vspace{-0.5cm}
\end{figure}

Inspired by the original arc current curve and the decomposed components of input signals, this paper proposes a simple and elegant activation function named Enhanced Adaptive Sine (EAS), which inherits the prior knowledge learned from input and Fourier decomposition 
analysis, and the excellent approximation ability in neural networks from super-expressive activation functions. The EAS is depicted in Fig~\ref{fig:An illustration of EAS} and is defined as:
\begin{equation*}\label{eq:EAS}
	\mathrm{EAS}(x)\coloneqq \begin{cases}
        \sin(\omega x+\varphi )  & \text{for} \  x\ge  0,\\
        \frac{x}{1+|x|}+\sin(\varphi) & \text{for} \  x<0,\\
	\end{cases}
\end{equation*}
{where the right part is a periodic ``Sine'' function with trainable parameters $\omega \in [0,\infty )$ and $\varphi \in [-\pi,\pi]$, and the left part is a real analytical asymptotic function.} In the asymptotic function $f$, its Taylor series at any point $x_0$ in its domain converges to $f(x)$ for $x$ in a neighborhood of $x_0$, point-wise. 
The proposed EAS can adaptively extract the key features in arcing signals by training $\omega$ and $\varphi$ since the input signal is a linear combination of several sub-signals with different frequencies in the frequency domain. This trait is beneficial for solving real-world problems. {Meanwhile, the super-expressiveness of the networks with EAS is related to the positive part of the activation function and is proven by~\cite{zhang2022deep}. Thus, it can help the model reduce the number of parameters.}


When EAS is applied in deep learning-based methods, the arc fault diagnosis problem can be interpreted as a multinomial distribution whose parameters $\theta$ are determined by fault labels, and these labels are converted to class probabilities by the softmax operator. For the $K$ classes arc fault diagnosis, the likelihood function for the tuple $(\bold X ,\bold Y )$ is given by:
\begin{equation} \label{standard classfication}
\begin{split}
&Pr(\bold Y|\bold X, \theta )= Softmax(f(\bold X, \theta))\\
\end{split}
\end{equation}
where $(\bold X ,\bold Y )$ are the input and output of the network $f(\cdot)$; $\theta $ is the parameters of the network. In $f(\cdot)$, the model can be described as:
\begin{equation} \label{network architecture using EAS}
\begin{split}
&f(\bold X, \theta) = \bold W_{n}(\phi_{n-1}\circ \phi_{n-2}\circ\cdots \circ \phi_{0})(\bold X)+\bold b_{n}\\
&x_{i}\longmapsto \phi_{i}(x_{i}) = EAS(\bold W_{i}x_{i} +\bold b_{i})
\end{split}
\end{equation}
where $\phi_{i}(\cdot)$ is the i-th layer of the network; $x_{i}$ is the input of i-th layer; $\bold W_{i}$ and $\bold b_{i}$ are the corresponding weight and bias. When EAS is applied in each layer, the model is forced to capture different phases and frequencies. The amplitude can be automatically adjusted by the weight of the next layer. If the frequency $\omega$ is constant, we can achieve the effect of extracting features of different frequencies by stacking more layers and activation functions, which has been demonstrated in image reconstruction. 

{The convergence of the proposed MFNN using EAS is tested in Fig~\ref{fig:The convergence of MFNN}. To emphasize the discrepancies in outcomes, we employ a logarithmic transformation ($\log$) during the visualization of loss function. Based on Fig~\ref{fig:The convergence of MFNN}, MFNN tends to be stable at about epoch=60, providing good performance in arc fault diagnosis.}


\subsection{Multi-frequency Neural Network Architecture}

\begin{figure*}[!t] 
\centering
\includegraphics[width=0.9\linewidth]{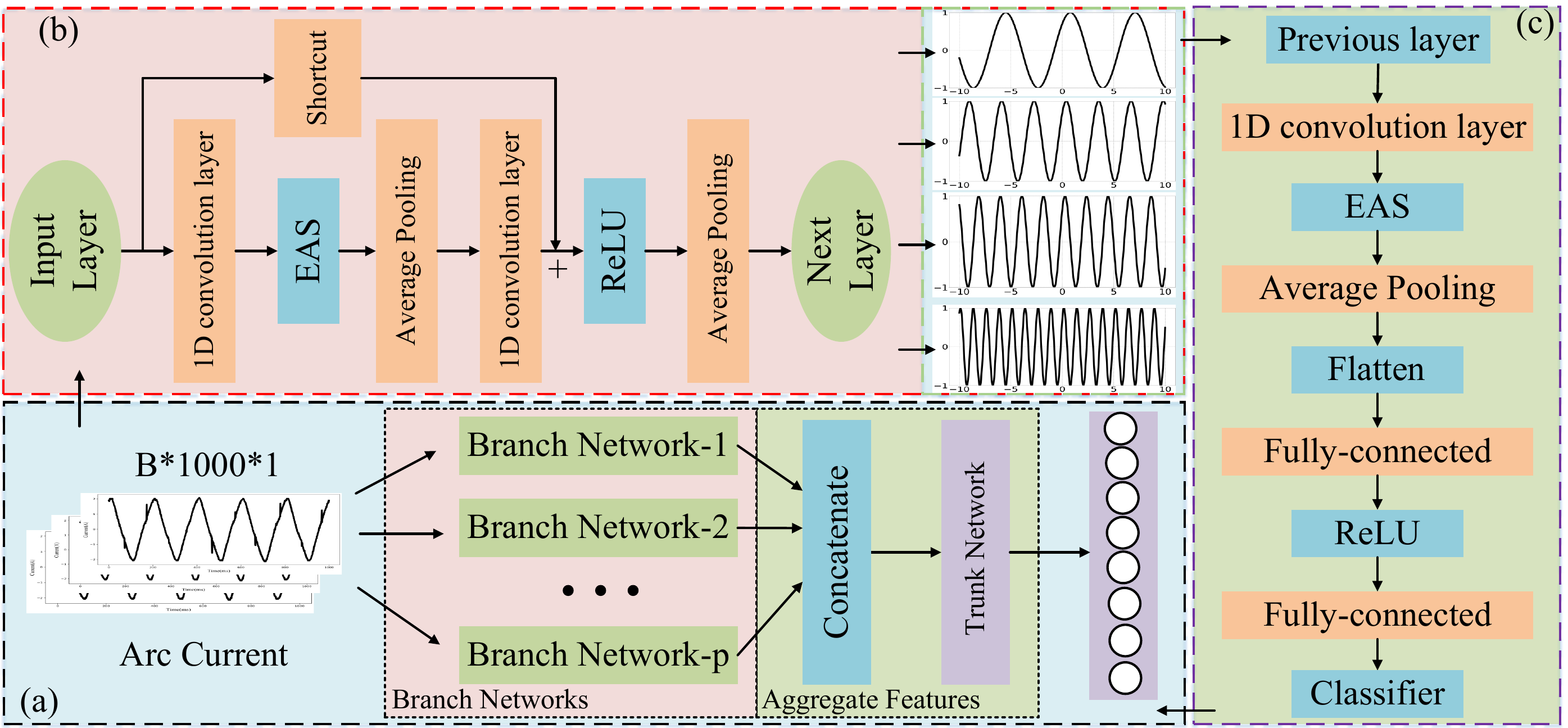}
\caption{Overview of the MFNN framework. (a) The flow chart of MFNN. (b) A branch network example. (c) The trunk network structure} 
\label{fig:Overview of the MFNN}
\vspace{-0.3cm}
\end{figure*}

The network architecture is inspired by the Discrete Fourier Transform, defining that the input signals are composed of multiple periodic sub-signals and can be captured by multiple branch networks. It is consistent with the idea of additive networks and the natural phenomenon of rivers converging. 
Simultaneously, this architecture is also used to learn nonlinear operators~\cite{lu2021learning} based on the universal approximation theorem of operators.

{Fig~\ref{fig:Overview of the MFNN} (a) depicts the network architecture of MFNN, including multiple parallel branch networks and a truck network, which are shown in Fig~\ref{fig:Overview of the MFNN} (b) and (c).} The branch networks are used to capture features with different frequencies and project them into high dimensions. Subsequently, these features are concatenated and extracted by the trunk network, which consists of a 1-D convolution layer, an EAS activation function, an average pooling layer, and multiple fully connected layers with ReLU. The arc fault diagnosis based on this architecture can be represented as 
\begin{equation} \label{standard classfication}
\begin{split}
& Branch_i(\bold X) = BranchNet_i(\bold X, \theta_i)\\
& AggregateF = Concat(\sum_{i=1}^{p} Branch_i(\bold X))\\
& Pr(\bold Y|\bold X, \theta ) = Softmax(TrunkNet(AggregateF))\\
\end{split}
\end{equation}

{Mathematically, we train the network by optimizing the following objective function:}
\begin{equation} \label{lab:loss}
\begin{split}
f^{*}=arg\min_{f} \frac{1}{N} \sum_{i=1}^{N}\ell(f(x_i,\theta),y_i)
\end{split}
\end{equation}
{where $\ell(\cdot)$ represents the cross entropy loss, $N$ is the number of samples in tuple $(\bold X ,\bold Y )$, and $f(\cdot)$ is MFNN, which is represented in Eq~\eqref{standard classfication}}.

This network architecture has an unexpected benefit, which is that it has strong noise resistance and can extract features from noisy data with fewer computational resources, thus improving the accuracy of arc fault diagnosis. The parallel branch network with EAS forcibly decomposes the input data into multi-frequency sub-signals. Among them, the noise signal will be represented by some branch networks, and the other branches can effectively represent the fault signal, since the representation of occasional fault signals is different from that of continuous noise. The contributions of the architecture network and the EAS activation function will be discussed in ablation experiments.


\subsection{Branch Network: Multi-Frequency Feature Extraction}

The branch network is designed to separately capture the local features of arcing current, especially the features with certain frequencies. By coupling the residual structure, the branch sub-model enables the network to capture broader features, which is equivalent to a deeper local feature network extractor.

Fig~\ref{fig:Overview of the MFNN} (b) gives an illustration of a branch network, explaining the information flow in the sub-model. The branch network work can be recognized as a high-frequency signal extractor, equipped with a periodic activation function. The extracted features will follow the 'Sine' pattern and have different frequencies and phases, helping MFNN locate the real fault.

Let $x\in \mathbb{R}^{B\times m \times n}$ be the input of the branch network and the output vector of i-th branch network is $Branch_i(x)\in \mathbb{R}^{B\times M \times N}$. The input is first extracted by the 1-D convolution layer to find the high-dimensional signal, and then EAS adds the nonlinearity and periodicity in the features, forcing the signal to be expressed as a sub-signal of the FT. 
We also notice that the arc fault features are usually a small area of distortion rather than an instantaneous mutation. Therefore, the average pooling is utilized in the branch network for on-linear down-sampling, reducing the spatial size of the representation and helping to find the mean feature. This part is the key to the branch network, expressed as:
\begin{equation} \label{Latent branch network}
\begin{split}
& Latent_{1}=f_{AvePool1}(EAS(f_{Conv1}(x)))
\end{split}
\end{equation}
where $Latent$ is the latent feature.

The residual structure is used in the branch network as well to provide global features for the trunk network. The subsequent activation function is set to ReLU instead of EAS to reduce the burden of back propagation, since half of the periodic signals can also reflect the whole periodic signals. Meanwhile, the stack of EAS can be equivalent to a single EAS with a higher frequency $\omega$, resulting in the model paying too much attention to the local higher frequency data, which is useless in arc fault diagnosis. This part can be given by
\begin{equation} \label{Branch_i}
\begin{split}
& Latent_2 = f_{Conv2}(Latent_1)+f_{ShortCut}(x) \\
& Branch_i(x)=f_{AvePool2}(ReLU(Latent2) \\
\end{split}
\end{equation}

\subsection{Trunk Network: Aggregated Features Analysis and Classification}

{The trunk network is used to analyze the features that come from the branch networks and are connected into a feature aggregation. Fig~\ref{fig:Overview of the MFNN} (c) gives an illustration of the trunk network. }

{Instead of summing the outputs of branch networks, we concatenate the extracted features for trunk network to diagnose. Specifically, aggregated features are analyzed using a 1D convolution layer, an EAS layer, and an average pooling layer to extract the potential detailed information from the global features, due to the limited number of branch networks. Then, all features are flattened for fully connected layers to classify specific arc faults. ReLU has the same contribution as the activation function in branch networks. The trunk network can be given by}
\begin{equation} \label{Trunk}
\begin{split}
& Latent_{trunk1}=f_{AvePool}(EAS(f_{Conv1}(x))) \\
& Latent_{trunk2}= ReLU(f_{Linear1}(Flatten(Latent_{trunk1}))) \\
& y=f_{Linear2}(Latent_{trunk2}) \\
\end{split}
\end{equation}

\section{Experimental Results}
\label{sec:Experiments}

{To further access the efficacy of the proposed neural network, we evaluate MFNN against a wide range of deep learning models, including AutoEncoder (AE)~\cite{9439848}, lightweight CNN (LCNN)~\cite{10018466}, ArcNet~\cite{9392282},  ArcNN~\cite{10054597}, Transformer~\cite{chabert2023transformer}, CNN-LSTM~\cite{10908361}, and BLS~\cite{10477286}, which have demonstrated their out-performance in arc faults diagnosis.} Table~\ref{Table: The PRM of all models} shows the number of parameters, flops and the peak memory access cost of the models, explaining the computation burden of each model. To compare these models, we conduct an experiment in the low-voltage electrical safety laboratory for electric appliances with/without arcing and analyze the corresponding current signals.


\subsection{Experiment Platform}

\begin{figure}[!t] 
\centering
\includegraphics[width=0.8\linewidth]{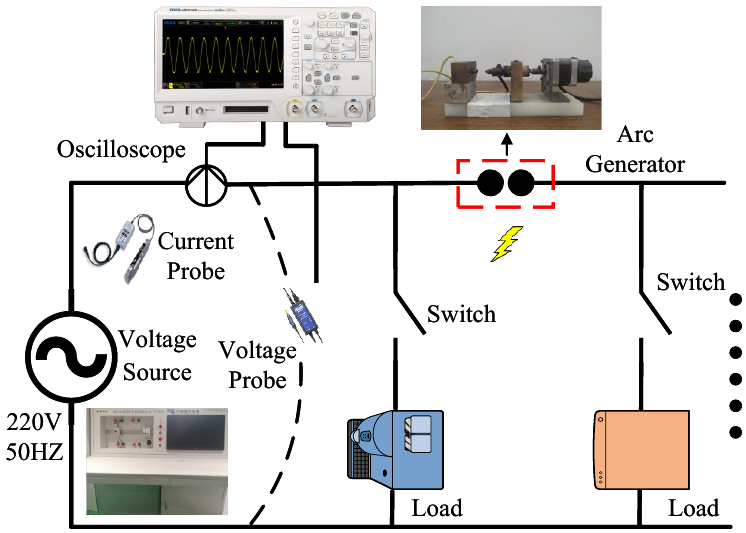}
\caption{Experimental platform of arc fault experiment.} 
\label{fig:Experimental platform}
\vspace{-0.5cm}
\end{figure}

The configuration of the experiment platform is shown in Fig~\ref{fig:Experimental platform}, including a voltage source, a voltage probe, a current probe, an oscilloscope, an arc generator, and loads. The voltage source works at 220V, 50Hz through the CKAT10 development kit. The parameters analyzed in this system (especially current and voltage) are captured by the Rigol oscilloscope MSO5000 (250 mA/S), CP8050A current probe (50A/50MHz), and the Rigol voltage probe (250 MHz). The arc is generated by an arc generator in accordance with GB 31143 and UL 1699 standards.

{The experiments are conducted on a variety of load types, including tungsten lamps (800W), heating kettles (1800W), air compressors (2200W), vacuum cleaners (1500W), switching power supplies (3.95A), dehumidifiers (178W), microwave ovens (1300W), and disinfection cabinets (800W). Arc fault datasets are obtained for individual loads or combined loads. Therefore, there are 8 tuples and 16 categories, including normal and fault currents. The arc gap can be flexibly adjusted between 0.5 mm and 3 mm to emulate various arc formation scenarios. The arc fault occurs at random time to simulate the real situation. All arcs are manually controlled to meet the effective arc duration requirement of less than 500 ms, ensuring the validity of detection experiments under standardized fault conditions.} 

In the experiments, there are ten experiments for each fault and each experiment contains 1,000,000 samples and two features (current and voltage).  The duration of each experiment is 5 seconds. 
In the dataset pre-processing, we split the normal data and arc fault data into multiple slices of length 10,000 (window = 10,000, step size = 5,000) and downsample the data. The dataset is discussed in Motivation~\ref{Motivations}. {To balance the classes, we keep the data of each class consistent.}
For the dataset diversity and the fairness of comparison, we perform low-resolution down-sampling and add different levels of noise to the dataset.

\begin{table}[!t]
\centering\footnotesize
\caption{The PRM, flops and MAC of main models. \#PRM denotes the number of parameters and \#MAC denotes the peak memory access cost.}
\label{Table: The PRM of all models}
\renewcommand{\arraystretch}{1.3}
\scalebox{1}{\begin{tabular}{ l| l| l| l }
\toprule
Models & \#PRM  & FLOPs & \#MAC   \\
\hline
AE & 32.75M  & 1049.15M &  2.01MB    \\
\hline
LCNN  & 1.04M  & 1098.66M & 3.25MB \\
\hline
ArcNet & 4.24M  & 5052.29M &  8.16MB  \\
\hline
ArcNN & 0.53M  & 776.3M &  2.05MB \\
\hline
Transformer & 8.21M  &  709.76M &  8.32MB  \\
\hline
CNN-LSTM &16.99M  & 8601.45M & 8.97MB  \\
\hline
BLS & 0.21M  & 6.68M &  0.32MB  \\
\hline
MFNN &  0.26M  & 29.24M &  1.41MB  \\
\bottomrule 
\end{tabular}}
\vspace{-0.3cm}
\end{table}

\subsection{Model Setup}

For each experiment, we train the models with a batch size of 32 using the `Adam' optimizer with an initialized learning rate of $0.001$. The learning rate decays by a factor of 0.5 every 30 epochs. The epoch is set to 100. The dataset is split into training, validation and test sets with a ratio of 0.8:0.1:0.1. {In MFNN, there are three branch networks. In each branch network, the arc fault currents are extracted by the 1D-convolution layers with 6 filters, and each kernel is $5\times 1$. The kernel size of the average pooling layer is 2. The shortcut is a 1D-convolution layer with the same parameters. In trunk network, the aggregated features are extracted by a 1D-convolution layer with 8 filters, and each kernel is $5\times 1$ as well. The fully connected layers have 256 neurons and ReLU as an activation function. All test accuracy are the average of five experiments.}
The experiments are implemented in 'Pytorch' using a CPU Intel i7-12700 processor at 2.10 GHz and a GPU NVIDIA GeForce RTX 4070Ti. 

\begin{table}[!t]
\centering\footnotesize
\caption{The test accuracy of arc fault diagnosis with different sampling times using a dataset with/without noise.}
\label{Table: Analytical Experiments with different sample time}
\renewcommand{\arraystretch}{1.3}
\scalebox{0.9}{\begin{tabular}{c| c| l l l l l }
\toprule
Dataset & Sample/$10^{-2}$ & 0.5 ms & 1 ms & 2.5 ms & 5 ms & 10 ms \\
\hline
\multirow{8}{*}{Noisy} & AE & 88.14\% &95.55\% & 98.65\% &99.44\% & 99.66\%  \\
 & LCNN & 94.44\% &98.88\% & 99.79\% &99.88\% & \textbf{100\%} \\
 & ArcNet & \textbf{96.66\%} &\textbf{99.07\%} & 99.47\% &99.55\% & 99.60\%  \\
 & ArcNN & 83.33\% &94.63\% & 98.28\% &99.33\% & 99.49\%  \\
 & Transformer & 89.56\% & 96.21\% & 98.88\% & 99.66\% &  99.72\%  \\
 & CNN-LSTM  & 91.01\% &  98.45\% &  \textbf{100\%} &  \textbf{100\%} &  \textbf{100\% } \\
 & BLS & 85.21\% & 90.45\% &  97.56\% & 98.21\% &  98.78\%  \\
 & MFNN & 90.00\% & 98.33\% & \textbf{100\%} &\textbf{100\%} & \textbf{100\%}  \\
\hline
\multirow{8}{*}{Normal} & AE &  95.55\% & 100\% &  100\%& 100\% &100\%  \\
 & LCNN & 96.66\% & 100\% &  100\%& 100\% & 100\% \\
 & ArcNet &  94.44\% & 100\% &  100\% &  100\% & 100\%  \\
 & ArcNN & 85.55\% & 97.22\% &  100\% &  100\% &100\%  \\
 & Transformer & 94.63\% & 99.89\% & 100\% & 100\% &  100\%  \\
 & CNN-LSTM  & 95.14\% &  99.99\% &  100\% &  100\% & 100\%  \\
 & BLS & 89.21\% & 97.45\% &  99.78\% & 100\% &  100\%  \\
 & MFNN & 95.56\%  &  99.98\% &  100\% &  100\% & 100\% \\
\bottomrule 
\end{tabular}}
\vspace{-0.3cm}
\end{table}

\subsection{Analytical Experiments}

The analytical experiment explores the efficiency of MFNN at different dataset resolutions. By choosing different sampling times, the dataset can be down-sampled with various resolutions.

Table~\ref{Table: Analytical Experiments with different sample time} summarizes the test accuracy of each model across different datasets with multiple sampling times. The Signal-to-Noise Ratio (SNR) of noisy data is set to 5. Before analyzing the details of the experiments, we notice an interesting phenomenon. In the two analysis experiments, it is not that the higher the data precision, the higher the accuracy. The most precise dataset cannot guarantee the highest test accuracy across the eight deep learning models. On the contrary, medium-precision data can achieve better performance.

For the normal dataset, which has little noise, MFNN has a comparative performance compared to other models, especially when the sampling time is higher than $1 \times 10^{-2}$ms. {Because the dataset is noise-free, it is hard to judge which model is better, except for the lowest test accuracy 85.55\% and 89.21\% from ArcNN and BLS. Whereas, when the noise is added to the dataset, we notice the under-performance of MFNN. When the sampling time $0.5\times 10^{-2}$ms, MFNN is No.4, achieving 90\% test accuracy, which is 6.66\%, 4.44\% and 1.01\% lower than top three models (ArcNet, LCNN, CNN-LSTM), respectively.} Nevertheless, as the sampling time gradually increases, the advantage of MFNN in test accuracy gradually emerges. It maintains 100\% accuracy when the sampling time transition from $2.5\times 10^{-2}$ms to $10\times 10^{-2}$ms. {Meanwhile, the test accuracy of other models is only around 99\%, except for the hybrid model CNN-LSTM. This experiment shows that in arc fault diagnosis, CNN-based models are more likely to find the necessary features for classification, improving the test accuracy.}

\begin{figure}[!t] 
\centering
\includegraphics[width=0.9\linewidth]{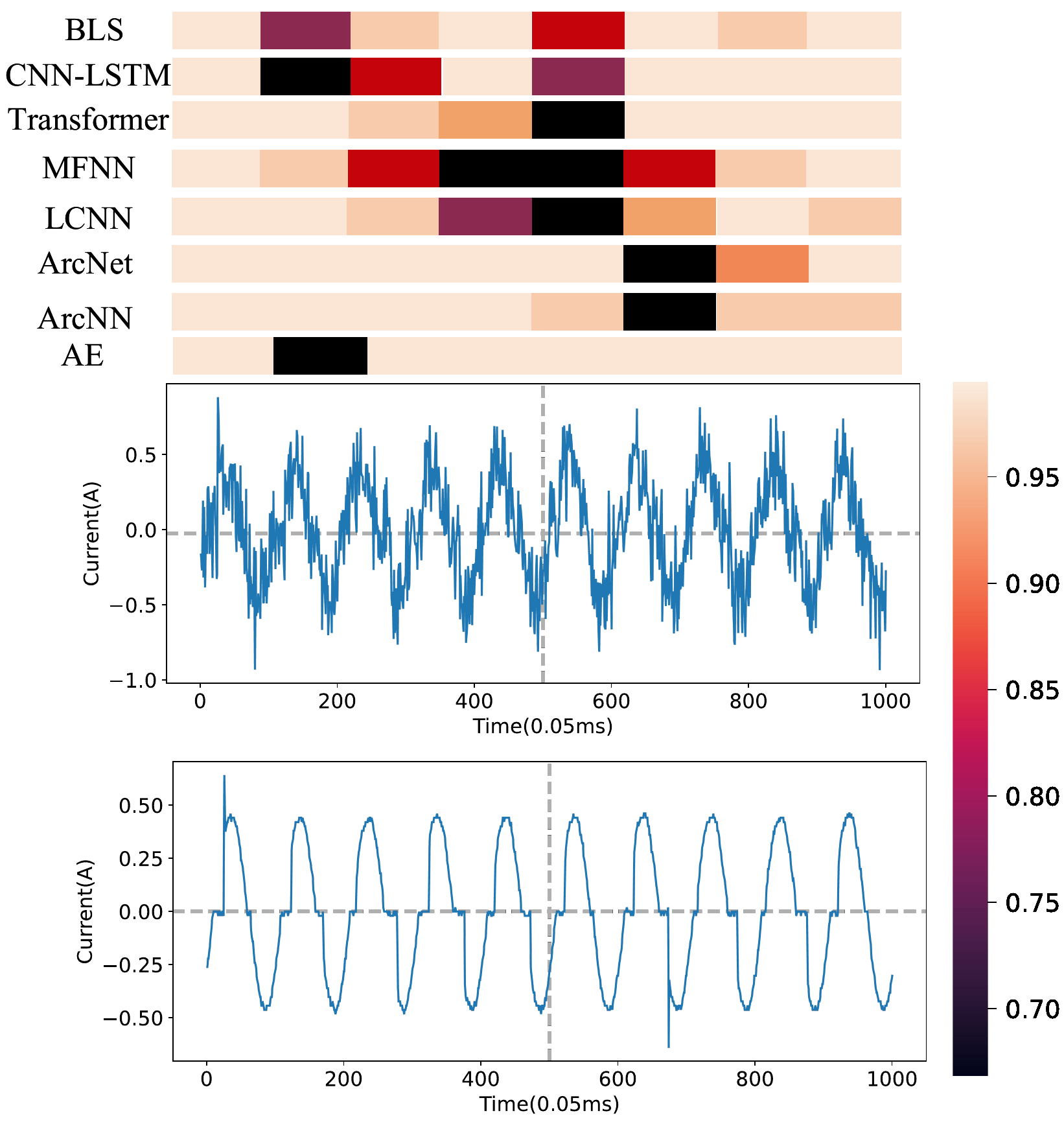}
\caption{Occlusion experiment visualization when sampling time is $5\times 10^{-2}$ms and SNR=5.} 
\label{fig:Occlusion Experiment of MFNN in Analytical Experiments}
\vspace{-0.5cm}
\end{figure}

To further evaluate the effectiveness and underline the arc fault location ability of MFNN, we conduct occlusion experiments at $SNR=5$ and the sampling time is $5\times 10^{-2}$ms, showing the extracted features of all models, since the test accuracy at this experiment is distinguishable. The experimental results are shown in Fig~\ref{fig:Occlusion Experiment of MFNN in Analytical Experiments}. The occluding sizes and strides are set to 200 and 100, respectively. The occluded pixels were all replaced by zeros. By replacing the occluded pixels, we can locate the features captured by the models through the change in test classification probability. The occlusion experiments explain why MFNN, LCNN and CNN-LSTM can outperform other arc fault diagnosis models. As deposited in the raw data, the arc faults occur at each periodic signal and are covered by the additional noise.
{ArcNet, ArcNN, AE, BLS and Transformer classify arc faults based on one main local feature, represented by the black or purple part in the figure.} The other “flat shoulder” features of the arcing current across the zero-crossing region are ignored, resulting in the potential degradation of generalization performance. In contrast to this, LCNN can find two main features in the arcing current, although one of them is not as important as the another one, represented as purple. {CNN-LSTM finds three features, and the influence of these three features gradually decreases, appearing in black, purple, and red, respectively.}
MFNN can locate four features, and two of them are the main features, giving more useful information to the model for generalization. The auxiliary features, which are colored red, can also be used to provide additional information for arc fault diagnosis.

\begin{figure}[!t]
\centering
\subfloat[Visualization of Branch Network-1]{
\includegraphics[width=0.85\linewidth]{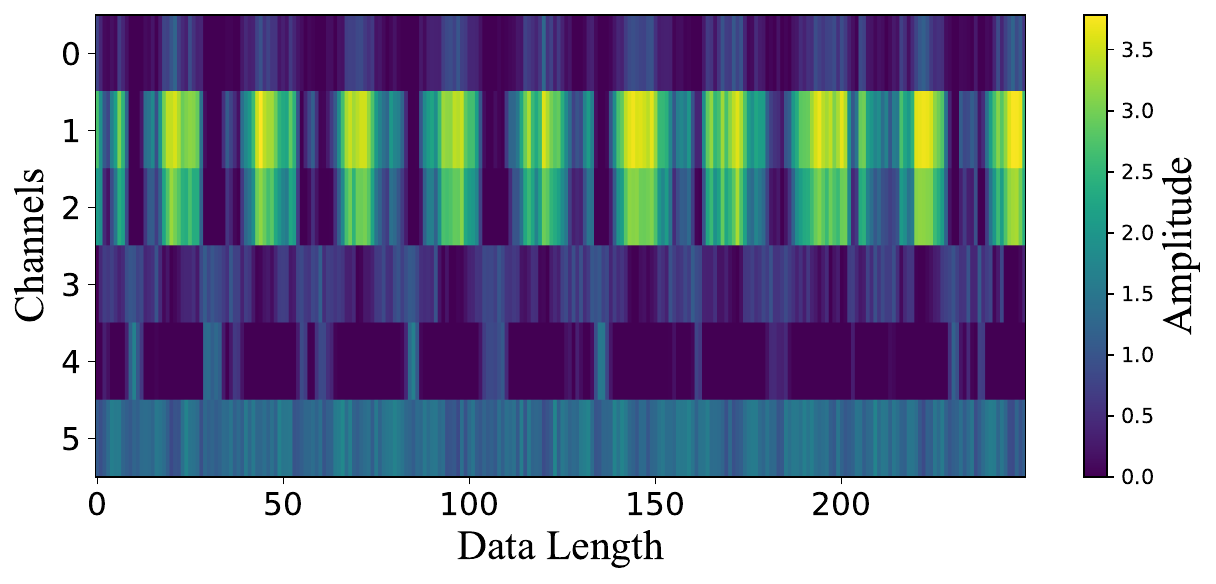}
\label{fig:branch1_heatmap}
} \\
\subfloat[Visualization of Branch Network-2]{
\includegraphics[width=0.85\linewidth]{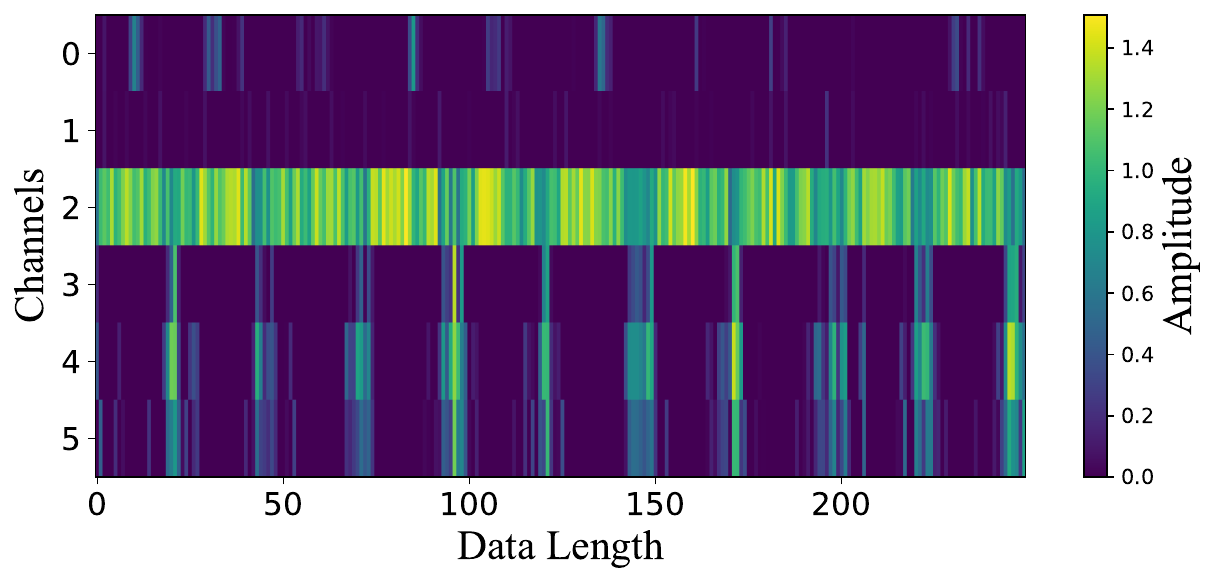}
\label{fig:branch2_heatmap}
} \\
\subfloat[Visualization of Branch Network-3]{
\includegraphics[width=0.85\linewidth]{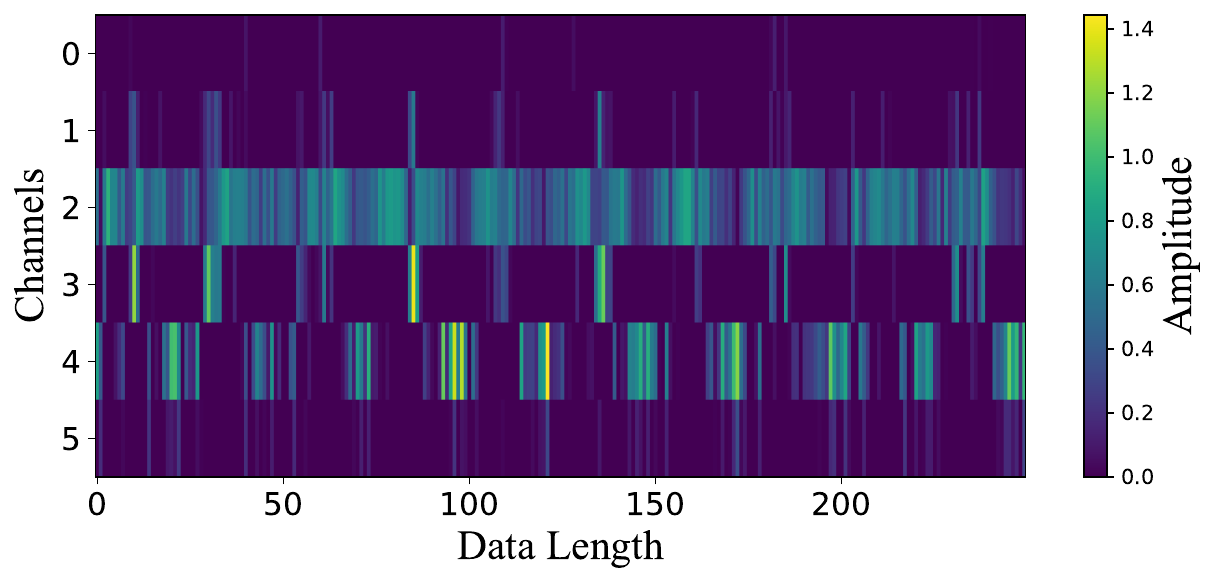}
\label{fig:branch3_heatmap}
} 
\caption{Visualization of Branch Networks of MFNN. }
\label{fig:Visualization of Branch Networks}
\vspace{-0.5cm}
\end{figure}

{To demonstrate that branch networks can effectively decompose the arcing current with different frequencies, helping MFNN find the arc faults, Fig~\ref{fig:Visualization of Branch Networks} visualizes the intermediate variables of MFNN, that is, the output of each branch network when the fault is same as Fig~\ref{fig:Occlusion Experiment of MFNN in Analytical Experiments}. Each branch network has 6 channels, and each channel shows the corresponding sub-signals. First, branch network 1 is responsible for the main signal decomposition, including low-frequency and high-amplitude sub-signals and medium-frequency and medium-amplitude sub-signals. 
Then, branch networks 2 and 3 can focus on locating the real arc fault (high-frequency with low-amplitude sub-signals) from noisy data, except for finding rest medium-frequency and medium-amplitude sub-signals in channel 2. We notice that the amplitudes of the channel 1 in branch network 2 and channel 0 in branch network 3 are almost zero. These signals can be the constant sub-signal and the `flat shoulder' signal, assisting MFNN diagnose arc faults. Other sub-signals can be recognized as the noise from experimental measurement and the added noise.}

\subsection{Comparative experiments}
To further test the robustness of MFNN, noise with different levels of SNR is added to the dataset for comparison. Meanwhile, to highlight the scalability of the proposed activation function, we try to use EAS in other models to achieve better performance.

\begin{table}[!t]
\centering\footnotesize
\caption{The test accuracy of arc fault diagnosis with different SNR.}
\label{Table: experiments with different SNR}
\renewcommand{\arraystretch}{1.3}
\scalebox{0.92}{\begin{tabular}{ l| l l l l l l l }
\toprule
Models  & -9 & -5 & -3& -1 & 1 & 3    \\
\hline
AE & 86.04\% &92.34\% & 96.17\% &96.35\% & 98.14\% & 98.70\%    \\
LCNN  & 82.03\% &94.43\% & 98.21\% & 99.47\% & 99.85\% & 99.88\%   \\
ArcNet & 88.50\% & 96.84\% & 98.77\% & 99.50\% & 99.52\% & 99.54\%  \\
ArcNN & 90.89\% &93.94\% & 97.32\% &98.36\% &98.44\% & 98.88\%  \\
Transformer & 86.71\% & 90.62\% & 94.19\% & 95.21\% &  97.76\%  &  99.33\%  \\
CNN-LSTM  & 91.91\% &  96.72\% &  98.88\% &  99.44\% & 99.65\%  &  99.88\% \\
BLS & 80.24\% & 84.52\% &  87.89\% & 93.69\% &  97.12\% &  98.06\%  \\
MFNN & \textbf{96.54\%} & \textbf{99.14\%}  &  \textbf{99.66\%} & \textbf{99.89\%} & \textbf{100\%} & \textbf{100\%}   \\
\bottomrule 
\end{tabular}}
\vspace{-0.3cm}
\end{table}

\subsubsection{Performance under different SNR} In field data, noise is inevitable, which is different from the dataset generated by experiments. In this part, we add additional noise to our dataset, setting the SNR to -9, -5, -1, 1, 3, respectively. The sampling time is $5\times 10^{-2}$ ms to ensure the appropriate dataset resolution and model performance distinction.

Table~\ref{Table: experiments with different SNR} compares the test accuracy of the models in the experimental arc fault datasets in multiple SNRs. Although all models achieve good performance at higher SNR, there are some subtle differences. {The noise immunity of BLS is obviously the worst among all models, because the test accuracy is only 93.69\% when SNR=-1, which is much lower than other models. This phenomenon can be attributed to fully connected layers, which are used for incremental learning with parameter reduction. The stability of Transformer-based arc fault diagnosis model is questionable because the test accuracy has been rapidly declining.} 
The performance of the rest models is stable, maintaining test accuracy within an acceptable range. However, due to the dilated convolution, AcrNN reduces the model parameters at the expense of a certain classification accuracy, compared to LCNN, ArcNet, and MFNN. Among all models, only MFNN can achieve 100\% test accuracy when SNR = 1 and 3, which is much better than other models.

When SNR is lower than -1, the difference in the model's performance in terms of noise immunity is greatly reflected in the table. There is a sudden drop in test accuracy in AE, LCNN, ArcNet, and CNN-LSTM when the SNR changes from -5 to -9, which means that when the noise level exceeds this limit, the model's performance drops rapidly. However, the limit of an ideal model should be as low as possible, so that it can maintain high accuracy and performance in the presence of a lot of noise. Benefiting from the dilated convolution, ArcNN's noise resistance has been improved. The larger receptive field ensures that ArcNN can discover more global features rather than being limited to local noise. {More parameters and more CNN stacking in CNN-LSTM can also improve the noise resistance to a certain extent.} The most impressive phenomenon is that the noise resistance performance of MFNN far exceeds that of other models. When SNR=-9, the test accuracy of MFNN can still reach 96.54\%, exceeding LCNN 14.51\% and BLS 16.3\%. Meanwhile, the test accuracy of the remaining experiments exceeded 99\%, which never occurs in other models. We also test that only when SNR=-15, the test accuracy of MFNN will be lower than 90\%, which is much better than others.

\subsubsection{The scalability of EAS} 
To verify the super-expressiveness, practicality, and scalability of EAS, we replaced some activation functions of AE, LCNN, ArcNet,  ArcNN, Transformer, CNN-LSTM, and BLS with EAS, based on the observation in the study~\cite{zhang2022deep} that only a few neurons with super-expressive activation functions are required to approximate functions with arbitrary precision, thus avoiding large generalization errors. 

Table~\ref{Table: experiments with combined activation functions} shows the test accuracy of the novel deep learning models with combined activation functions. {Overall, the combination of EAS and ReLU can improve the accuracy of the model, especially when the effect of the original model itself is not ideal, for example, AE, ArcNN and BLS.} However, the combination is not a panacea. When the SNR is high and the model can make full use of the data, the improvement in test accuracy is limited or may even be worse than the original model. The outstanding performance of the hybrid activation function is concentrated in the data with a lot of noise. The maximum accuracy improvement is 4.47\%, when AE uses the hybrid activation function in the dataset with SNR=-9. The performance of other models with a hybrid activation function is impressive as well, as the test accuracy always improves when SNR$\le -3$. Therefore, EAS is practical and scalable in deep learning-based arc fault diagnosis, especially when a large amount of noise is mixed into the data.

\begin{table}[!t]
\centering\footnotesize
\caption{The test accuracy of arc fault diagnosis with combined activation functions.}
\label{Table: experiments with combined activation functions}
\renewcommand{\arraystretch}{1.3}
\scalebox{0.92}{\begin{tabular}{ l| l l l l l l l }
\toprule
Models  & -9 & -5 & -3& -1 & 1 & 3    \\
\hline
AE & 90.51\% & 92.74\% &  98.21\% &  98.32\% & 98.44\% &  98.77\%    \\
LCNN  & 83.25\% & 96.76\% & 98.32 \% &  99.33\% &  99.55\% &  99.79\%   \\
ArcNet & 91.51\% & \textbf{97.54\%} & \textbf{98.99\%} & \textbf{99.44\%} &  \textbf{99.66\%} &  99.66\%  \\
ArcNN & \textbf{94.08\%} & 95.53\% & 98.88\% & \textbf{99.44\% }& \textbf{99.66\%} &  \textbf{99.77\%}  \\
Transformer & 87.37\% & 91.69\% & 94.58\% & 95.19\% &  97.88\%  &  99.39\%  \\
CNN-LSTM  & 92.45\% &  96.78\% &  98.89\% &  99.41\% & 99.59\%  &  99.82\% \\
BLS & 81.11\% & 85.67\% &  87.92\% & 94.31\% &  97.65\% &  98.03\%  \\
\bottomrule 
\end{tabular}}
\vspace{-0.3cm}
\end{table}

\subsection{Ablation Experiments}
In this subsection, we explore the necessity of branch networks and EAS 
in MFNN. By designing an additional end-to-end model with only one trunk network (1-Trunk), which has a similar parameter number, we can demonstrate that the branch networks, inspired by the Fourier Transform, contribute to the out-performance of MFNN. Meanwhile, we replace the activation functions of MFNN with ReLU to verify the necessity of EAS (ReLU-M).

Since the advantage of MFNN is noise resistance, we conduct the experiments using datasets with SNR=-9, -5, -1, 1, 3 respectively, and the sampling time is $5\times10^{-2}$ ms. 
Table~\ref{Table: experiments of trunk model across datasets with different SNR} shows the test accuracy of all ablation experiments

\subsubsection{The necessity of branch networks}

\begin{table}[!t]
\centering\footnotesize
\caption{The test accuracy of Ablation Experiments across datasets with different SNR.}
\label{Table: experiments of trunk model across datasets with different SNR}
\renewcommand{\arraystretch}{1.3}
\scalebox{0.95}{\begin{tabular}{ l| l l l l l l l }
\toprule
Models  & -9 & -5 & -3& -1 & 1 & 3    \\
\hline
1-Trunk & 94.42\% & 97.21\% &  98.88\% &  98.54\% & 99.88\% & 100\%    \\
ReLU-M & 96.09\% & 98.88\% &  99.44\% &  99.78\% & 100\% & 100\%    \\
\bottomrule 
\end{tabular}}
\vspace{-0.3cm}
\end{table}

The first line of Table~\ref{Table: experiments of trunk model across datasets with different SNR} summarizes the test accuracy of the trunk model across datasets with different SNR. First, the well-designed trunk network outperforms the conventionally designed network, which is shown in Table~\ref{Table: experiments with different SNR}. Compared with the models with hybrid activation functions in Table~\ref{Table: experiments with combined activation functions}, it achieves comparative or better performance, especially when SNR=-9. Whereas, the test accuracy drops significantly when we replace the branch networks with trunk networks. This might be because the branch networks can decompose the original data into sub-signals with multiple frequencies to capture local fault features instead of repeatedly extracting from global features. Therefore, the Fourier Transform-inspired network architecture can contribute to the arc fault diagnosis, improving the accuracy of classification.

\subsubsection{The necessity of EAS} 

The second line of Table~\ref{Table: experiments of trunk model across datasets with different SNR} summarizes the test accuracy of MFNN with only ReLU across datasets with different SNR. Different from 1-Trunk, the test accuracy just has a slight decrease, about 0.1\%~0.4\%, compared to the proposed MFNN. It remains a comparative performance because branch networks with ReLU can be recognized as polynomial functions, which can also be used to decompose signals. However, the efficiency is inferior to that of the proposed MFNN if the input signals are too complex. In general, the branch networks and EAS in MFNN both contribute to the arc fault diagnosis, and the branch network plays a major role.


\subsection{Hardware Implementation of MFNN}
\begin{figure}[!t]
\centering
\subfloat[Hardware Implementation of MFNN.]{
\includegraphics[width=0.4\linewidth]{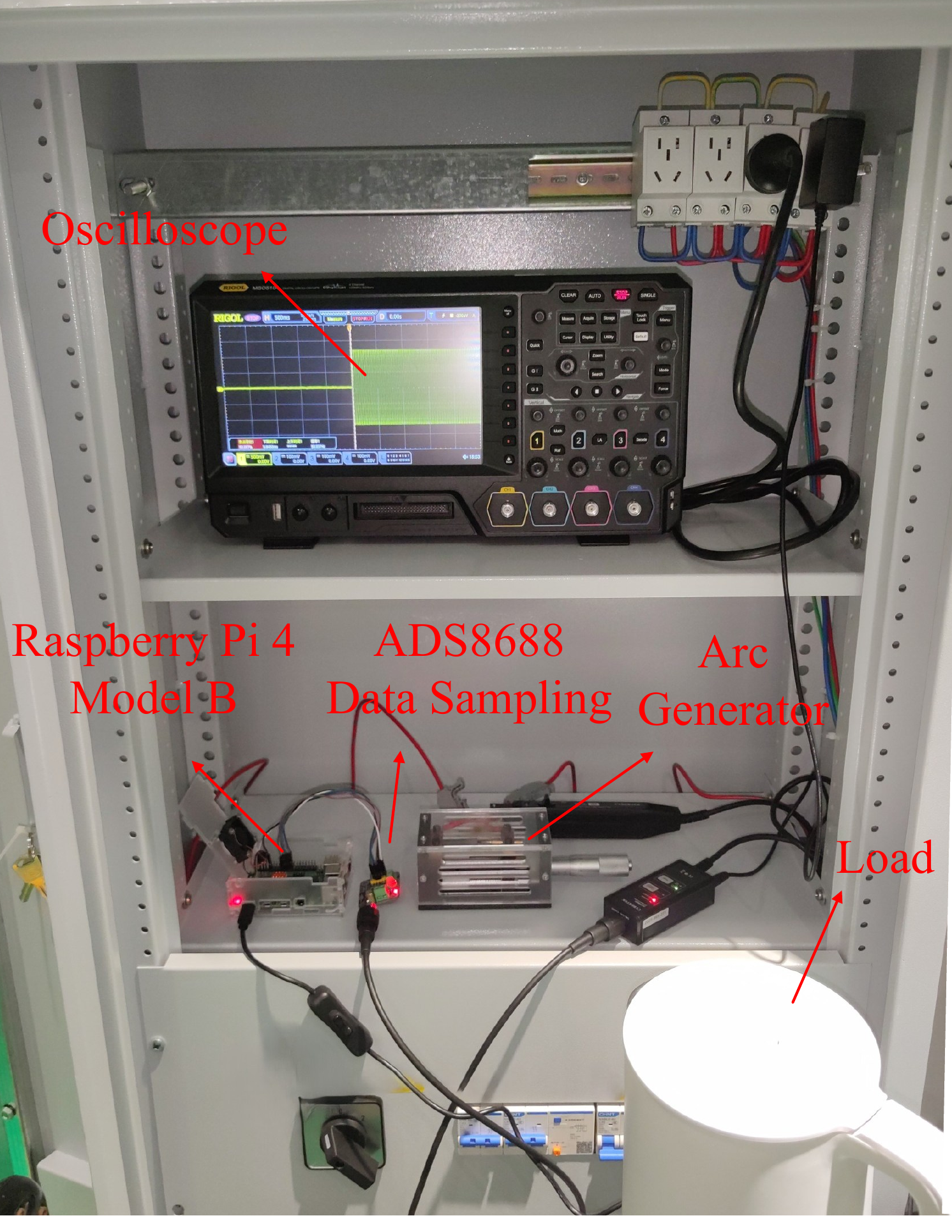}
\label{fig:Hardware Implementation of MFNN}
} ~
\subfloat[Confusion matrix of MFNN.]{
\includegraphics[width=0.5\linewidth]{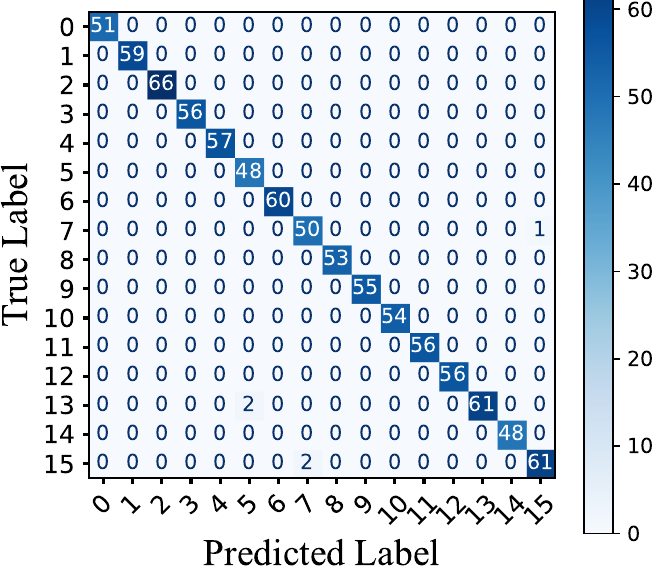}
\label{fig:Confusion matrix of MFNN}
} \\
\caption{Hardware implementation and results. }
\label{Hardware implementation and results}
\vspace{-0.5cm}
\end{figure}

{To test the practicality of MFNN in arc fault diagnosis, we embed MFNN in a Raspberry Pi 4B, shown in Fig~\ref{fig:Hardware Implementation of MFNN}. The main devices are the ADS8688 data acquisition module and the Raspberry Pi 4B. The ADS8688 data acquisition module integrates a 16-bit successive approximation register analog-to-digital converter and operates at a maximum sampling rate of 1 MSPS, supporting bipolar input ranges and features built-in over-voltage protection.
The Raspberry Pi 4B consists of an ARM Cortex-A72 1.5GHz CPU processor, a 4GB LPDDR4, and a 500MHZ broadcom VideoCore vi GPU. It also has HDMI, USB3.0 and other ports.}

{The arcing current is captured by the current probe and digitized by the ADS8688 data acquisition module. The collected data is transmitted to a Raspberry Pi 4B via a serial peripheral interface. MFNN deployed on the Raspberry Pi processes the acquired current waveform to distinguish between fault and normal currents. We test the hardware implementation experiments for 896 times and the average runtime of MFNN is 4.2 ms, satisfying the IEC62606 standard. Fig~\ref{fig:Confusion matrix of MFNN} shows the confusion matrix of MFNN. As depicted in the figure, in the hardware implementation experiments, MFNN has almost no mis-classification and meets the detection accuracy requirements.}

\section{Conclusion}\label{sec:Conclusion}

Deep learning models have achieved outstanding performance in arc fault diagnosis. Whereas, the inherent problem in arc fault diagnosis is that without physical guidance, whether these highly accurate data-driven models are trustworthy. In this light, this paper is dedicated to proposing a novel and elegant neural network named MFNN, inspired by prior physical knowledge, by introducing a super-expressive activation function and a well-designed network architecture within limited computational resources. eight advanced deep learning based arc fault diagnosis models are compared on datasets with multiple sampling times and multi-level noise to verify the efficacy of MFNN. It is demonstrated to be superior to other networks in analytical experiments. Meanwhile, the comparative experiments and hardware experiments demonstrate the noise immunity of MFNN, the practicality, and scalability of EAS. The contributions of the proposed network architecture to arc fault diagnosis are discussed in ablation experiments.

{The key limitation of MFNN lies in the prior knowledge of the input, which greatly affects the performance of networks.  
The two proposed components, EAS and network architecture, are based on prior knowledge of the arcing current and the corresponding analysis methods, especially the periodicity and signal decomposition.} That is to say, this method holds the potential for adaptation to other domains within industrial diagnostic applications, if the input dataset has similar physical knowledge. {The following research will focus on prior knowledge extension and domain adaptation in arc fault diagnosis.}


\end{document}